\documentclass[a4paper]{article}
\usepackage{amsmath, amssymb}
\usepackage{amsfonts}
\usepackage{multicol}
\usepackage{braket}

\usepackage[dvipdfmx]{graphicx}
\usepackage{ascmac}

\usepackage{fancybox}
\usepackage{cite}
\usepackage{amsthm}

\usepackage{color}

\newcommand{\eq}[1]{\begin{equation} #1 \end{equation}}
\newcommand{\eqa}[2]{\begin{equation} #1 \label{#2} \end{equation}}
\newcommand{\balign}[1]{\begin{align} #1 \end{align}}

\newcommand{\fn}{\footnote}

\newcommand{\todayd}{\the\year/\the\month/\the\day}

\newcommand{\bib}{\bibitem}

\newcommand{\eref}[1]{Eq.~\eqref{#1}}

\newcommand{\bel}{\begin{easylist}}
\newcommand{\eel}{\end{easylist}}
\def \({\left(}
\def \){\right)}
\def \[{\left[}
\def \]{\right]}
\newcommand{\la}{\left\langle}
\newcommand{\ra}{\right\rangle}
\newcommand{\sumtwo}[2]%
{\mathop{\sum_{#1}}_{#2}}
\newcommand{\sumthree}[3]%
{\mathop{\mathop{\sum_{#1}}_{#2}}_{#3}}
\newcommand{\sumfour}[4]%
{\mathop{\mathop{\mathop{\sum_{#1}}_{#2}}_{#3}}_{#4}} 
\newcommand{\prodtwo}[2]%
{\mathop{\prod_{#1}}_{#2}}
\newcommand{\mintwo}[2]%
{\mathop{\min_{#1}}_{#2}}
\newcommand{\maxtwo}[2]%
{\mathop{\max_{#1}}_{#2}}
\newcommand{\maxthree}[3]%
{\mathop{\mathop{\max_{#1}}_{#2}}_{#3}}
\newcommand{\limtwo}[2]%
{\mathop{\lim_{#1}}_{#2}}
\newcommand{\suptwo}[2]%
{\mathop{\sup_{#1}}_{#2}}
\newcommand{\supthree}[3]%
{\mathop{\mathop{\sup_{#1}}_{#2}}_{#3}}
\newcommand{\supfour}[4]%
{\mathop{\mathop{\mathop{\sup_{#1}}_{#2}}_{#3}}_{#4}} 
\newcommand{\inftwo}[2]%
{\mathop{\inf_{#1}}_{#2}}
\newcommand{\infthree}[3]%
{\mathop{\mathop{\inf_{#1}}_{#2}}_{#3}}
\newcommand{\inffour}[4]%
{\mathop{\mathop{\mathop{\inf_{#1}}_{#2}}_{#3}}_{#4}} 
\def\rnum#1{\resizebox{0.5em}{\height}{\expandafter{\romannumeral #1}}}
\def\Rnum#1{\resizebox{0.5em}{\height}{\uppercase\expandafter{\romannumeral #1}}}

\makeatletter
  \newcommand{\subsubsubsection}{\@startsection{paragraph}{4}{\z@}%
    {1.0\Cvs \@plus.5\Cdp \@minus.2\Cdp}%
    {.1\Cvs \@plus.3\Cdp}%
    {\reset@font\sffamily\normalsize}
  }
\makeatother
\makeatletter
\def\verbatim@font{\normalfont\fontfamily{txr}\selectfont
\let\do\do@noligs
\verbatim@nolig@list}
\makeatother
\begin{document}

\noindent \textbf{\Large  Finite-time thermodynamic uncertainty relation do not hold for discrete-time Markov process} \hfill 

\begin{flushright}
\textbf{Naoto Shiraishi}\fn{
Department of physics, Keio university, Hiyoshi, Kohoku-ku, Yokohama, 223-8522, Japan
}
\end{flushright}

\begin{abstract}
Discrete-time counterpart of thermodynamic uncertainty relation (conjectured in P. Pietzonka, {\it et.al.},arXiv:1702.07699 (2017)) with finite time interval is considered.
We show that this relation do not hold by constructing a concrete counterexample to this.
Our finding suggests that the proof of thermodynamic uncertainty relation with finite time interval, if true, should strongly rely on the fact that the time is continuous.
\end{abstract}

\section{Introduction}

Recently, a trade-off relation between fluctuation of current and entropy production in stationary Markov processes, called {\it thermodynamic uncertainty relation}, has been investigated intensively~\cite{BS15, PBS16, Jordan, Jordan-full, SST16, PRS17, PNRJ, Maes17, HH17}.
The relation for the case of long-time limit for stationary Markov jump processes (Markov processes with continuous time) on discrete states was first conjectured~\cite{BS15}, and then proved by using the large-deviation technique~\cite{Jordan, Jordan-full}.
The relation was also proved for the case of short-time limit~\cite{SST16}, in a completely different context, for a broad class of Markov processes.
On the other hand, the thermodynamic uncertainty relation for the case of finite time interval is proved only for a specific situation~\cite{PNRJ} and its general form is only conjectured~\cite{PRS17}.

Although the original conjecture on this relation is on continuous-time Markov processes, in this note we consider a discrete-time counterpart (i.e., the case of Markov chain).
We shall show that the discrete-time counterpart do not hold by constructing a concrete counterexample to the relation.

\section{Thermodynamic uncertainty relation}

We first explain the claim of the thermodynamic uncertainty relation conjectured in Ref.~\cite{PRS17}.
In the following, we normalize $kT=1$.

Consider a Markov jump process on discrete states in the non-equilibrium steady state.
We assume that there is no parity-odd variable and field.
Let $a_{ij}$ be an empirical current with the transition from the state $i$ to the state $j$ in time interval $0\leq \tau\leq t$, which counts how many jumps from $i$ to $j$ occur in $0\leq \tau\leq t$.
A generalized time-integrated current $X$ is defined as
\eq{
X:=\sum_{i,j}d_{ij}a_{ij}
}
with coefficients satisfying $d_{ij}=-d_{ji}$.

By denoting the average of this stochastic process and the transition rate from $i$ to $j$ by $\la \cdot \ra$ and $k_{ij}$ respectively, the average current $J$, the variance of time-integrated current ${\rm Var}[X]$, the entropy production rate $\sigma$ are expressed as
\balign{
J&:=\frac{\la X\ra}{t} \\
{\rm Var}[X]&:=\la X^2\ra -\la X\ra ^2 \\
\sigma&:=\sum_{i,j}P_i^{\rm ss}k_{ij} \ln \frac{P_i^{\rm ss}k_{ij}}{P_j^{\rm ss}k_{ji}},
}
where $P_i^{\rm ss}$ represents the stationary distribution of the state $i$.
The thermodynamic uncertainty relation claims the following inequality:
\eqa{
\frac{{\rm Var}[X]\sigma}{J^2t}\geq 2.
}{tur-jump}
This inequality implies that to realize large current $J$ large fluctuation of current or much dissipation inevitably accompanies.

\section{Construction of counterexample}

The discrete-time counterpart of \eref{tur-jump} is obtained by replacing the time interval $t$ to the number of steps and the transition rate to the transition probability.
We here, however, demonstrate that this counterpart is not valid.

Consider a system with two states, $A$ and $B$.
There are two transition paths, 1 and 2, between $A$ and $B$
\fn{
We note that it is irrelevant to our discussion that there are two transition paths between the same pair of states.
In fact, our discussion is fully valid for a Markov chain on four states $A$, $B$, $C$, and $D$, and the transition paths $AB$ and $CD$ ($BC$ and $DA$) have the same property as that with transition 1 (2).
}.
The transition probability of the transition 1 is set to
\balign{
k_{AB}^1=&\frac{2}{y} \\
k_{BA}^1=&\frac{1}{y} ,
}
and that of the transition 2 is 
\balign{
k_{AB}^2=&\frac{1}{y} \\
k_{BA}^2=&\frac{2}{y}.
}
Here, $y\geq 3$ is a parameter which we will fix later.
The staying probability $k_{AA}$ and $k_{BB}$ is then given by $(y-3)/y$.
The stationary distribution is $P_A^{\rm ss}=P_B^{\rm ss}=1/2$.

The average entropy production per step is given by
\eq{
\sigma =\frac{1}{y}\ln 2.
}
We consider current of transition 1 from $A$ to $B$, and set this current to $X$ (i.e., $d_{AB}^1=-d_{BA}^1=1$ and $d_{AB}^2=d_{BA}^2=0$).
The average current is given by
\eq{
J=\frac{\la X(t)\ra}{t}=\frac{1}{2y}.
}

We shall show that the discrete time counterpart of the uncertainty relation \eqref{tur-jump} is not satisfied for the case with $t=2$ and proper choice of $y$.
In two steps, $X$ can take 1,0, or $-1$.
The probability of each event is calculated as
\balign{
P(X=1)&=\frac{2(y-1)}{y^2} \\
P(X=0)&=\frac{y^2-3y+4}{y^2} \\
P(X=-1)&=\frac{y-2}{y^2}.
}
Hence, the variance of $X$ is given by
\eq{
{\rm Var}[X]=\frac{3y-5}{y^2}.
}
The left-hand side of the thermodynamic uncertainty relation \eqref{tur-jump} is then calculated as
\eq{
\frac{{\rm Var}[X]\sigma}{J^2t}=\( 3-\frac{5}{y}\) 2\ln 2.
}
Because the right-hand side of the thermodynamic uncertainty relation is 2, the relation is violated if $y$ satisfies
\eq{
3\leq y\leq \frac{5}{3-1/\ln 2}=3.21067\cdots .
}

\section{Discussion}

We remark that our finding does not suggest the incorrectness of the conjectured thermodynamic uncertainty relation for Markov jump processes.
Instead, our finding clarifies the fact that we cannot prove the original uncertainty relation (if true) by using a method which is also applicable to Markov chain.
For example, an approach based on the fluctuation theorem will not lead to the proof of the thermodynamic uncertainty relation, because the fluctuation theorem and techniques used in the proof of it are valid for both Markov jump processes and Markov chain.
To prove this, inherent properties to continuous-time Markov processes are inevitable.

\section*{Acknowledgment}

The author thanks Shin-ichi Sasa and Sosuke Ito for helpful comments on this manuscript.
This work was supported by Grant-in-Aid for JSPS Fellows JP17J00393.

\end{document}